# Real-time observation of magnetization and magnon dynamics in a two-dimensional topological antiferromagnet MnBi$_2$Te$_4$

F. Michael Bartram[1,2,#], Meng Li[1,#], Liangyang Liu[1], Zhiming Xu[1], Yongchao Wang[3], Mengqian Che[1], Hao Li[4], Yang Wu[4,5], Yong Xu[1,6,7,8], Jinsong Zhang[1,6,7], Shuo Yang[1,6,7] and Luyi Yang[1,2,6,7,*]

[1]State Key Laboratory of Low Dimensional Quantum Physics, Department of Physics, Tsinghua University, Beijing 100084, China

[2]Department of Physics, University of Toronto, Toronto, Ontario M5S 1A7, Canada

[3]Beijing Innovation Center for Future Chips, Tsinghua University, Beijing 100084, China

[4]Tsinghua-Foxconn Nanotechnology Research Center, Department of Physics, Tsinghua University, Beijing 100084, China

[5]Beijing Univ Chem Technol, Coll Sci, Beijing, 100029, China

[6]Frontier Science Center for Quantum Information, Beijing 100084, China

[7]Collaborative Innovation Center of Quantum Matter, Beijing 100084, China

[8]RIKEN Center for Emergent Matter Science (CEMS), Wako, Saitama 351-0198, Japan

[#]These authors contributed equally.

[*]e-mail: luyi-yang@mail.tsinghua.edu.cn

## Abstract

Atomically thin van der Waals magnetic materials have not only provided a fertile playground to explore basic physics in the two-dimensional (2D) limit but also created vast opportunities for novel ultrafast functional devices. Here we systematically investigate ultrafast magnetization dynamics and spin wave dynamics in few-layer topological antiferromagnetic MnBi$_2$Te$_4$ crystals as a function of layer number, temperature, and magnetic field. We find laser-induced (de)magnetization processes can be used to accurately track the distinct magnetic states in different magnetic field regimes, including showing clear odd-even layer number effects. In addition, strongly field-dependent antiferromagnetic magnon modes with tens of gigahertz frequencies are optically generated and directly observed in the time domain. Remarkably, we find that magnetization and magnon dynamics can be observed in not only the time-resolved magneto-optical Kerr effect but also the time resolved reflectivity, indicating strong correlation between the magnetic state and electronic structure. These measurements present the first comprehensive overview of ultrafast spin dynamics in this novel 2D antiferromagnet, paving the way for potential applications in 2D antiferromagnetic spintronics and magnonics as well as further studies of ultrafast control of both magnetization and topological quantum states.

## 1. Introduction

Ultrafast pump-probe techniques have proved to be a powerful tool for understanding the dynamics of magnetic materials on short timescales, such as demagnetization, excitation of collective modes [1]. Femtosecond laser pulses can also be used to control magnetic order, including exciting transient magnetic states inaccessible in the static regime [1-3]. Although most ultrafast research has been performed in ferromagnets, these techniques have also demonstrated their power in the probing and manipulation of spins in antiferromagnets [2], which are otherwise hard to detect (especially for small samples) due to their vanishing net magnetic moment.

The discovery of 2D magnetic materials about 6 years ago [4-11] has provided new opportunities for both fundamental studies of spin physics in reduced dimensions and for the creation of novel ultrafast communication, logic and memory devices based on 2D van der Waals magnets and their heterostructures. Unfortunately, due to their small volume the magnetic order in atomically thin magnetic materials cannot be probed by conventional tools such as neutron scattering. Using optical techniques based on magneto-optical effects has turned out to be a very effective alternative and has been widely applied to determine magnetic states in 2D magnetic materials [6,7,11]. However, to date most work has focused on the static magnetic order; ultrafast spin dynamics – which are both fundamentally important and crucial for ultrafast device applications – have received much less attention and remain largely unexplored.

As the first intrinsic magnetic topological insulator [16-22], atomically thin $MnBi_2Te_4$ crystals bridge three fascinating realms in condensed matter physics: magnetism, topology and 2D materials. Below a critical temperature of ~24 K, $MnBi_2Te_4$ forms an A-type antiferromagnetic order, with intralayer ferromagnetic coupling and antiferromagnetic coupling between adjacent layers. The properties of few-layer $MnBi_2Te_4$ have been probed by magneto-optical circular dichroism [23,24] and magnetic force microscopy [25], wherein the much stronger intralayer coupling makes each layer act as a single unit, allowing few-layer $MnBi_2Te_4$ to be modeled as a finite one-dimensional antiferromagnetic chain. $MnBi_2Te_4$ further exhibits quantum anomalous Hall and axion insulator phases for odd- and even-layer samples respectively [21,22], demonstrating the connection between magnetism and topology [26-29]. Ultrafast magnetization dynamics have been reported in bulk $MnBi_2Te_4$ [30], and a recent magneto-Raman study reported the detection of magnon modes in bilayer $MnBi_2Te_4$ [31]. Recent theoretical and experimental studies have also demonstrated optical axion electrodynamics and optical control of the antiferromagnetic order in even layer $MnBi_2Te_4$ [32,33]. However, systematic studies of ultrafast spin dynamics in atomically thin crystals of $MnBi_2Te_4$ (or in 2D magnetic materials in general) are still at an early stage [14,15], as these studies are technically demanding, involving ultrafast time-resolved spectroscopy combined with high-spatial-resolution microscopy, cryogenic temperatures and high magnetic fields. Sufficient understanding of ultrafast (de)magnetization mechanisms, timescales and magnon dynamics is critical for the development of 2D ultrafast data processing and storage devices.

Here we explore, for the first time, ultrafast dynamics of magnetization and magnons in 4- to 8-septuple layer (SL) thick $MnBi_2Te_4$ samples via measurements of both the time-resolved magneto-optical Kerr effect (MOKE) and the time-resolved reflectivity. While time-resolved MOKE is well-established probe of magnetization dynamics, we also observe a large magnetization-dependent reflectivity signal in our samples, implying an unusually strong connection between electronic states and magnetic order in this material. The magnetic state of few-layer $MnBi_2Te_4$ under an applied magnetic field differs distinctly depending on whether the number of layers is even or odd, due to an odd layer of numbers creating a necessary asymmetry between AFM sublattices (e.g., $N$+1 up, $N$ down for 2$N$+1 layers) [23,24]. We have observed laser-induced (de)magnetization processes which are highly sensitive to the magnetic state, which reflect these odd-even layer effects and also allow us to accurately track magnetic transitions across varying applied magnetic fields. In fact, this use of pump-probe techniques for probing the static magnetic state turns out to be a more effective tool compared to a conventional static method, which could be of use in studying other 2D magnets. In addition, by applying an in-plane external field we directly observe collective spin wave excitations in real time and show that their frequency and damping time also depend on both the even-or-odd layer number and the external magnetic field. The magnetization and recovery timescales and magnon frequencies and lifetimes revealed in this work provide critical parameters for further designing 2D ultrafast spintronic/magnonic devices based on few-layer MnBi$_2$Te$_4$ and heterostructures. Our results also highlight some apparent limitations of the current theoretical models, suggesting some need for improvements in the understanding of the magnetic interactions. Since topological quantum states in MnBi$_2$Te$_4$ are closely linked to the magnetic states, it is important to have a clear picture of the magnetic behavior. By presenting a comprehensive overview of pump-induced magnetization dynamics, we can therefore provide a basis for future experiments investigating ultrafast manipulation of topological states.

## 2. Methods

MnBi$_2$Te$_4$ crystals were prepared by mechanical exfoliation as described previously [22,34]. The layer thickness was determined by optical contrast, atomic force microscopy and time-resolved reflectivity measurements of the interlayer breathing mode [35].

To probe the nonequilibrium magnetization and spin wave dynamics, we performed two-color pump-probe reflectivity and MOKE measurements with pulsed lasers in two different configurations (see Supplementary Note 1 for details). In the first case, the magnetic field was applied perpendicular to the sample plane (Faraday geometry, Fig. 1a). In the second case, the magnetic field was almost in the sample plane, but with a small tilting angle of ~ 7.5 degrees (close to the Voigt geometry, Fig. 5a). In both cases, the pump and probe beams were at normal incidence, where the probe is sensitive to the change of the out-of-plane magnetization.

## 3. Ultrafast magnetization dynamics

We start with experiments performed in the Faraday geometry (shown schematically in Fig. 1a), where the external field is out-of-plane. Previous reports have studied the magnetic states at varying fields in such a configuration using static MOKE measurements [23,24]. As in those reports, our static MOKE measurements (shown in Fig. 1c,d for 4- and 5-layer samples at 3 K) show large steps at the transition from antiferromagnetic (AFM) order to canted antiferromagnetic (cAFM) order, as indicated by the lower edge of the shaded regions (~2 T and ~3.5 T for the 4- and 5-layer sample respectively). At a higher field, the cAFM order eventually coalesces in ferromagnetic (FM) order, with all spins aligned to the external field, indicated by the upper edge of the shaded regions (~3.3 T and ~6 T for the 4- and 5-layer samples). In static MOKE measurements, the cAFM-FM transition is marked by the signal flattening out, a change which can often be challenging to clearly identify [23,24]. As we will show, this transition can be more clearly distinguished by time-resolved measurements. We also note that like previous studies, we observe an anomalous hysteresis loop in the AFM region of even-layered samples. The exact origin of this loop remains under debate, which could result from uncompensated magnetization [23] or from optical axion electrodynamic response [32] as recently observed experimentally [33]. We do not attempt to address it here as it is not the focus of this work but note that it consistently appears in both our static MOKE and time-resolved data.

Figure 1b shows the time-resolved Kerr rotation for a 4-layer region of the sample at 3 K with an external field of 7 T, which is typical of the signals in this configuration. The signal is negative (corresponding to a decrease in magnetization) and shows four distinct components. First, there is a step-like feature at $t$ = 0 ($t_0$), followed by a slow rise, with the signal reaching the maximum (negative) value after about ~1 ns. Part of the signal then recovers on a timescale of a few nanoseconds, with the remainder lasting on a scale which we estimate to be ~100 ns, much longer than the interval of the adjacent pump pulses (12.5 ns), causing a large background offset (even before $t_0$) due to accumulation from previous pump pulses. We also note that this time-resolved response shows a strong non-linearity with respect to the pump and probe beam powers due to laser heating effects (we show a clear example of this for the second configuration in Supplementary Figure 6). We therefore have minimized the powers of the pump and probe beams as much as possible while maintaining acceptable signal quality.

This two-step demagnetization signal is very similar to typical cases of ultrafast demagnetization in rare-earth ferromagnets (e.g., Gd) [36] or ferromagnetic half-metals or insulators (e.g., La$_{1-x}$Sr$_x$MnO$_3$, CoCr$_2$S$_4$, CrI$_3$) [37,38]. Reports on bulk MnBi$_2$Te$_4$ samples have also shown similar results [30]. Such cases are generally described by the microscopic three-temperature model [36], which we discuss in more detail in Supplementary Note 2. The pump deposits energy into electronic states, which become very hot before rapidly cooling back down within a few ps as they thermalize with the lattice. During this hot electron phase, rapid demagnetization occurs, giving the step-like feature. In the next phase, the magnetization slowly shifts towards the appropriate equilibrium value for the new (increased) temperature of the system, leading to a slow rise time. We note that the ~ns scale rise time in these measurements is relatively long compared to that of many other materials (e.g., Ni, Gd) [1,36,37], but comparable to that of half-metallic ferromagnets (e.g., La$_{1-x}$Sr$_x$MnO$_3$) [37]. The long demagnetization time in MnBi$_2$Te$_4$ may be because the Mn $d$ bands are far away from the Fermi level [18] and therefore the Elliot-Yafet type spin relaxation is limited (see Supplementary Note 2 for further discussion). Finally, the signal will slowly decay as the entire system cools back to the ambient temperature via thermal diffusion. In our case there are two distinct decay times, one relatively short (~ a few ns) and one very long (on the order of 100 ns), which can be explained as originating from the out-of-plane diffusion (into the SiO$_2$ substrate) and in-plane diffusion (within the sample) respectively. For comparison, a very rough estimate of the in-plane diffusion timescale can be given by $d^2/D_{ip}$, where $d$ is the pump spot diameter and $D_{ip}$ is the in-plane thermal diffusivity. Using a diffusivity of ~0.073 μm$^2$ ns$^{-1}$ (from specific heat and in-plane thermal conductivity values in [39,40]) and a pump spot diameter of 2 μm gives a similar timescale of ~50 ns.

A simple measure of the field-dependent behavior of the signal can be obtained via the long-lived background, which we measure using the value of the signal just before $t_0$ (corresponding to ~12 ns after the previous pulse, where all faster components have already decayed away). Sweeping the external field back and forth gives results shown in Fig. 1 e,f for 4SL and 5SL samples at 3 K, which are antisymmetric about the magnetic field and show obviously layer-dependent patterns, including large peaks at the AFM-cAFM transition, and a noticeable change in the signal at the cAFM-FM transition. These features match extremely well with the features in the static MOKE signals and are notable in that they are easier to identify, particularly in the case of the cAFM-FM transition. We additionally include static MOKE data and background sweeps for 6 to 8 SL in Supplementary Figure 1, with similar results. By taking temperature dependent background sweeps for the 5-layer sample, shown in Supplementary Figure 2, we can also see how the transition fields evolve with temperature. These results demonstrate the efficacy of pump-probe measurements as a tool for probing the static magnetic order of few-layer MnBi$_2$Te$_4$, which could also be widely applied to the studies of other atomically thin magnetic materials.

Based on our three-temperature model, this long-lived background corresponds to an elevated sample temperature from pumping, meaning we are effectively measuring the change in $\theta_K$ with temperature. To verify this, we have also measured the static MOKE signals at 10 K, where we plot the difference between the 10K and 3K data in Fig. 1g,h. This plot shows clear similarities to the background sweep, suggesting that thermal effects are indeed the main driver of that signal. We also note that these signals are independent of the pump polarization, further supporting the conclusion of thermally driven transient magnetization dynamics [1].

Since typically, the polar Kerr rotation is proportional to the out-of-plane magnetization, we also calculate $\Delta M_z/\Delta T$ from our theoretical model (discussed in detail in Supplementary Note 3) for comparison, shown in Figure 2a and 2b. The results are qualitatively very similar, with peaks near the AFM-cAFM transitions and a distinct cusp at the cAFM-FM transition. In the AFM region, $\Delta M_z/\Delta T$ in the 4-layer region starts from zero (at zero field) and rises as the field increases, in contrast to the 5-layer region, in which it starts at a negative value and decreases in amplitude towards zero as the field increases. This odd-even layer effect exactly mirrors what is seen in the data, aside from a small offset in the 4-layer region due to the anomalous hysteresis loop. One notable difference is that the cAFM-FM transition is predicted to be much larger by theory compared to what we observe in the experiment. We show in Figure 2c transition fields for a variety of thicknesses extracted from our data (crosses) compared to theoretical calculations (dots), where while good agreement can be found for the AFM-cAFM transition points ($H_{cAFM}$), the cAFM-FM transitions ($H_{FM}$) observed in the data show an unexpected rapid decrease for thinner samples. This issue has not been noted in previous studies [23,24], where the focus was on the (more obvious) AFM-cAFM transition points.

Full time- and field- dependent data for 7- and 8-layer samples are shown in Figure 3. The overall behavior roughly just tracks that of the background sweeps, with peaks in amplitude near the AFM-cAFM transition points and a small additional peak at the transition between two cAFM configurations [23] in the 8-layer sample. One difference of note is that in the even-layered AFM region at low field, where the background is positive (due to magnetization increasing with temperature in this region, as discussed above), the initial step-like feature remains negative, meaning it still represents a decrease in magnetization. We can qualitatively understand this as interactions in the hot electron phase causing large demagnetizations in a small fraction of spins, compared to the later interactions causing smaller but more widespread shifts towards the new equilibrium (which has a larger magnetization, in the case of the even-layered sample). As a uniquely AFM behavior, such a feature is not well captured by the standard three-temperature model, which is designed to describe ferromagnetic systems. We also note that the timescales change somewhat as a function of field, though within the FM region they appear (nearly) constant. However, detailed quantitative analysis of changes in dynamics for different applied fields proves difficult, likely due to the influence of the various transition points. Around the AFM-cAFM transitions, we were sometimes able to observe small oscillatory signals (see Supplementary Figure 3), which are linked to the magnon modes discussed later. Measuring a set of different thicknesses all in the FM state suggested a trend of increasing rise (and fall) times for thicker samples (shown in Supplementary Figure 4), which implies a significant contribution to the underlying interactions from the interface.

Interestingly, in addition to the time-resolved Kerr signal, we can also observe a large time-resolved reflectivity change at low temperature, shown in Figure 4. Like the Kerr data, we observe clear features near the transition points, with the noticeable difference of being (approximately) symmetric with respect to the external field (b,c). We have verified that the signal also shows a sudden drop in amplitude above the Neel temperature (a inset). Such pronounced magnetorefractive effects do not exist in many other magnetic materials, suggesting that the electronic structure in MnBi$_2$Te$_4$ is unusually sensitive to the magnetic order. To further support this idea, we have performed first-principles DFT calculations (Supplementary Note 5) which indeed show a noticeable difference in calculated reflectivity between FM and AFM states. Surprisingly, a large signal can be observed in an even-layered sample with no external field, where the net magnetic moment (and any changes induced by heating) should be very small (ideally zero). The fact that it is relatively large in the zero-field even-layered samples suggests that these reflectivity changes cannot simply be thought of in terms of changes in net magnetic moment (as is conventionally done in MOKE), and might instead be related to changes in the Neel vector [2] or some more novel effect from the axion insulator state [32,33]. Separate reports have also indicated measurable shifts in the static reflectivity as bulk samples are cooled below the transition temperature [41], lending further experimental support to our observations.

## 4. Coherent spin wave dynamics

Next, we discuss the measurements in the (near) Voigt geometry, where the external field direction is set at a small ~7.5-degree angle to the sample plane, as shown in Fig 3a. In this configuration, any non-zero applied field causes the magnetic moments to begin tilting, forming a cAFM state, with the spins eventually aligning with the field, giving FM order. The tilt of the magnetic field slightly out of the sample plane gives us better control of the zero-field magnetic state and leads to magnon modes which cause modulations in the out-of-plane magnetization, which is typically necessary for detection.

Figure 5b displays a set of time-resolved reflectivity data for various applied magnetic fields in a 5-layer sample at 4K, where prominent oscillations with tens of GHz frequency starting right after $t_0$ emerge at around 3 T before again vanishing near 7 T. We note that observing magnons with time-resolved reflectivity measurements is unusual, though it has also been done in other materials such as EuTe [42] (via magnetization-induced bandgap shifts) and CrSBr [15] (via coupling between magnons and excitons). As discussed above, we have observed a strong magnetorefractive effect in this system, so the fact that the magnon signal also appears is not surprising. In fact, the oscillations are present in both reflectivity and Kerr angle, but the signal to noise ratio turns out to be much better in the reflectivity measurements (shown in Supplementary Figure 5). As before, the sample is highly sensitive to laser heating, which can be demonstrated here by observing that increasing the power causes a noticeable shift in the oscillation frequency (Supplementary Figure 6). We emphasize that time-resolved techniques are capable of detecting dynamics in a wide frequency range from sub-GHz to a few THz, making them particularly suitable for probing the ~20-50 GHz frequency magnons observed here. By contrast, Raman spectroscopy is limited by the low energy cutoff (usually around 3 cm$^{-1}$ ≈ 90 GHz) and traditional ferromagnetic resonance techniques are limited to lower frequencies (typically below ~40 GHz).

To show the full field dependence of this oscillatory signal, we plot the FFT amplitude across field and frequency in Fig. 5c. The frequency clearly decreases towards zero as the field approaches ~6.8 T, after which the signal vanishes. As before, we observe a long-lived (thermal) component of the time-resolved signals, which can be used to probe the magnetic state. In the background sweep of the time-resolved Kerr rotation (shown in Fig 5d) a large cusp can be seen at the same ~6.8 T field value where the magnons vanish, which we identify as the cAFM-FM transition field. A comparison of background sweeps across different thickness is shown in Supplementary Figure 7, along with theoretical calculations of $\Delta M_z/\Delta T$. Surprisingly, such a calculation does not resemble the observed data, which may be explained by including sensitivity to the Neel vector, similar to what was done in studies of the anomalous central loop in even-layered samples [32,33]. We have additionally measured the 5-layer sample at a selection of higher temperatures (shown in Supplementary Figure 8), wherein the magnon frequencies decrease with increasing temperature and the field where the magnon signal drops to zero (which decreases at higher temperature) consistently matches with the cusp in the corresponding background sweep. For a theoretical comparison, we plot the field dependence of magnon frequencies calculated via a linearized Landau-Lifshitz-Gilbert (LLG) equation (detailed in Supplementary Note 4) in Fig. 5e, wherein the lowest frequency mode just below the cAFM-FM transition (highlighted region) shows an obvious similarity to our measurements, also dropping to zero frequency at the transition from cAFM to FM order.

In addition to the 5-layer sample, we also observe similar oscillations in other regions of the sample (varying between 4 and 8 layers thick), which we show in Fig. 6 (all at 4 K), along with parameters obtained from fitting to a damped sinusoid plus a slow exponentially decaying background. The FFT amplitude is shown as a function of field and frequency, with the fitted oscillation frequencies marked with green dots (a-e), alongside the initial amplitude of the oscillation from the fit (f,g) and the oscillation lifetime (h,i). A clear difference between the even- and odd-layered samples can be observed. The even-layered samples show visible oscillations down to generally lower fields than in the odd-layered ones, and redshift slightly at the lowest fields. For all the even-layered samples the lifetime also steadily increases with decreasing field, whereas in the odd-layered samples the lifetime reaches a maximum and then decreases. The increase of spin wave damping towards the saturation field is likely due to the inhomogeneous broadening of local magnetic fields from doping or strain, in line with previous work in bilayer $CrI_3$ [14]. For comparison, we calculate the damping parameter $\alpha = 1/(f_0 \tau)$ from the frequency $f_0$ and lifetime $\tau$ (shown in Supplementary Figure 9) where we find that our damping parameter is somewhat larger (0.1 – 10) than what was observed observed in $CrI_3$ (<0.4).

Extrapolating our measurements towards zero field indicates a finite spin wave gap which slightly decreases from ~60 GHz (~0.25 meV) for thicker samples down to ~40 GHz (~0.17 meV) for the thinnest region, consistent with that observed in bilayer $MnBi_2Te_4$ by magneto-Raman measurements (~0.2 meV) [31]. In the theoretical calculation, the spin-wave gap would be expected to be layer-independent for even-layered samples, given by $\gamma\sqrt{K(K+2J)}$, with slightly reduced gap for thinner odd-layered samples (down to the obvious result of $\gamma K$ for monolayer), where $\gamma$ is the gyromagnetic ratio. We note that as mentioned before, the thinner samples show a reduced cAFM-FM transition field not well captured by our model, which may be related to why the frequency decreases despite theoretically being layer-independent. Using our estimated values of $J = 2.3$ T and $K = 0.35J$ with the gyromagnetic ratio of a free electron ($\frac{\gamma}{2\pi}$~28 GHz T$^{-1}$) gives value of 58 GHz for the even-layered spin-wave gap, which matches well with the observed data on the thicker regions. For a bulk sample, eliminating the influence of the surface via periodic boundary conditions gives a slightly increased gap of $\gamma\sqrt{K(K+4J)}$ = 79 GHz (0.33 meV).

This is somewhat smaller than the value of 0.4-0.5 meV obtained from inelastic neutron scattering measurements on bulk samples [43], but this may be simply due to sample-dependent variations in the anisotropy, as in some previously reported MOKE measurements [23,24] the observed AFM-cAFM transition field has been higher, suggesting a higher anisotropy level of $K \sim 0.6\,J$ (alongside a slightly higher $J$ value of 2.55 T), which would in turn lead to a higher bulk spin-wave gap of about 116 GHz (0.48 meV).

In principle, an *N*-layer system holds *N* magnon modes, but as mentioned above, the only mode observed here appears to be the lowest order mode, which is an "out-of-phase" mode wherein each spin precesses directly out of phase with its neighbors (see Supplementary Note 4 for further description of this mode). The signal is only visible above an onset field of a few tesla (which is smaller in even-layered samples) and vanishes again above the cAFM-FM transition. Note that the previous report on magnons in bilayer MnBi$_2$Te$_4$ using Raman spectroscopy observed the higher order of the two modes [31], in which the spins all precess in phase. To try to explain the limited field range in which magnons can be observed, we have performed theoretical calculations (shown in detail in Supplementary Note 4) which suggest that modeling the effect of the pump pulse as suddenly changing the effective exchange coupling gives a qualitatively similar result, with no excitation occurring in the FM region and a falling amplitude at low fields, with the odd-layered samples falling off more rapidly. A similar light-induced change in exchange coupling was also proposed for the excitation of the out-of-phase magnon mode in EuTe [42]. Note that another angle that can be taken is to consider the detection sensitivity. In a typical time-resolved MOKE experiment, the signal is taken to be proportional to modulations in the net out-of-plane magnetization $M_z$ caused by precession of the spins. As mentioned previously, this was part of the motivation for tilting the field slightly out-of-plane, since a purely in-plane field would result in this magnon mode having no effect on $M_z$ for even-layered samples. Even with the field tilted out-of-plane, this remains true in the FM region, which serves as an alternate explanation for the lack of signal above $H_{FM}$. However, the effects on $M_z$ are nonzero throughout the cAFM state for both odd- and even-layered samples, even increasing somewhat at lower fields, so this cannot serve as an explanation of the low-field cutoff observed in our data. In addition, the detection of magnons via time-resolved reflectivity makes it difficult to draw clear conclusions in this way, as it is unclear whether signal amplitude should even be related to changes in $M_z$.

## 5. Conclusion

In summary, we have observed magnetization and magnon dynamics in atomically thin $MnBi_2Te_4$ across a range of thicknesses. We have found that the magnetization-dependent signals are observed not only in time-resolved MOKE but also in time-resolved reflectivity, indicating strong coupling between the magnetic state and electronic structure. The time-resolved MOKE (and reflectivity) signals can accurately track different magnetic regimes, which could also be applied to the studies of other 2D van der Waals antiferromagnets. Here, this revealed a lower-than-expected transition into FM order for thin samples, which is difficult to detect via static MOKE measurements. We have also demonstrated direct optical excitation and detection of coherent magnons, whose frequencies can be tuned by external magnetic fields. Our work helps establish the magnetic response of $MnBi_2Te_4$ to femtosecond pulses in the few-layer limit, giving a basis for future work on optical control of spins or topological states [1-3,32,33,44,45]. The magnon frequencies in few-layer $MnBi_2Te_4$ are compatible with those of existing qubit technologies, showing that it has the potential for magnonic applications [46,47] in 2D devices. Many of the magnon dynamics discussed here are widely applicable to layered antiferromagnetic systems, and similar observations have been reported in previously mentioned materials like $CrI_3$ [14] and CrSBr [15]. Some techniques used in these studies, such as voltage gating the sample or measuring propagation of magnons through the sample, could be applicable to future studies of $MnBi_2Te_4$. We hope this work can also help inform studies of other layered antiferromagnets in the future, which may benefit from the techniques and analysis used in our experiments.

**Competing interests**

The authors declare no competing interests.

## Acknowledgements

Sample preparation, ultrafast optical measurements and calculations were all carried out at Tsinghua University. The work was supported by the National Key R&D Program of China (Grant Nos. 2020YFA0308800 and 2021YFA1400100), the National Natural Science Foundation of China (Grant No. 12074212). Y.Wu was supported by the National Natural Science Foundation of China (Grant No. 21975140 and 51991343) and Fundamental Research Funds for the Central Universities (Buctrc202212). F.M.B. was also supported by funds from the University of Toronto.

## Author contributions

Luyi Yang conceived and supervised the project.  F. Michael Bartram built the static MOKE and time-resolved MOKE and reflectivity experiments and performed all the optical measurements with help from Liangyang Liu and Mengqian Che supervised by Luyi Yang.  F. Michael Bartram carried out all the data analysis. Meng Li and F. Michael Bartram performed the theoretical calculations supervised by Suo Yang and Luyi Yang.  Zhiming Xu. performed the DFT calculations supervised by Yong Xu.  Hao Li and Yang Wu grew the samples.  Yongchao Wang fabricated the few-layer samples supervised by Jinsong Zhang.  Yong Xu and Shuo Yang provided theoretical insights.  F. Michael Bartram, Meng Li, Zhiming Xu and Luyi Yang wrote the paper in consultation with all authors.

## Appendix A. Supplementary materials

Supplementary materials to this article can be found online at

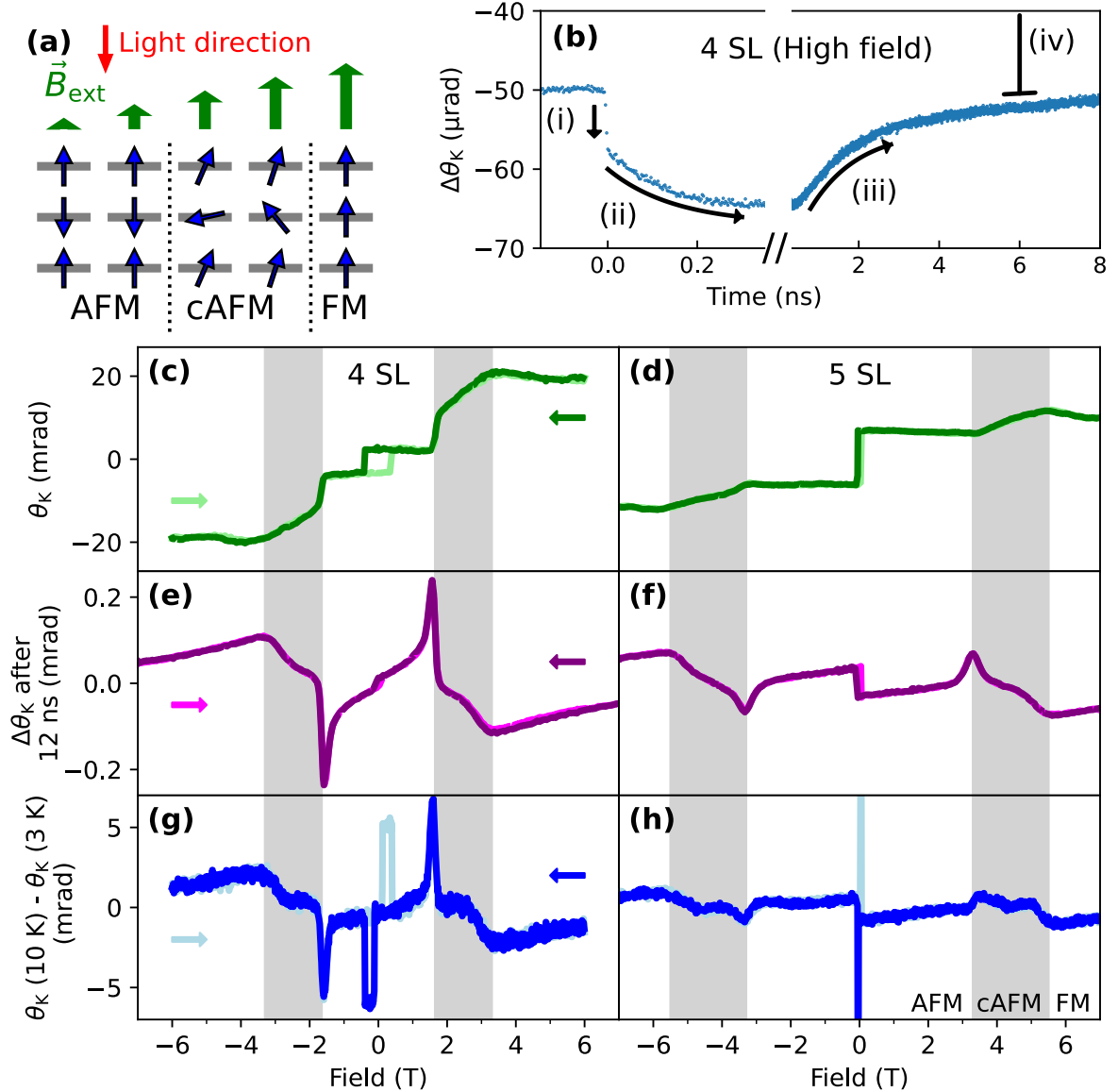

**Figure 1.** (Color online) **Static and time-resolved Kerr rotation with out-of-plane external field.** We apply a field in the out-of-plane direction, shown schematically in **a**, wherein the sample transitions from AFM to cAFM to FM order as the field strength increases. An example of a typical signal at 3 K is shown in **b**, where we observed 4 distinct features: (i) a sudden change immediately after the pump, (ii) a slow ns-scale rise, (iii) a few-ns scale decay, and (iv) a much larger and longer-lived signal that decays on a scale of ~100 ns. This latter feature can be easily measured by the signal just before the pump (~12ns after the previous pump pulse), where we can observe a striking field-dependence that varies with sample thicknesses (shown in **e** and **f** for 4 and 5 SL respectively). We compare this to static MOKE measurements done with a 633 nm HeNe laser (**c,d**) on the same sample regions (also at 3 K). In **g,h** we show the difference between static MOKE measurements at 10 K and 3 K, which show very similar features to those seen in the time-resolved measurements, suggesting a thermally driven mechanism. Arrows in **c**, **e**, **g** indicate the field sweep direction.

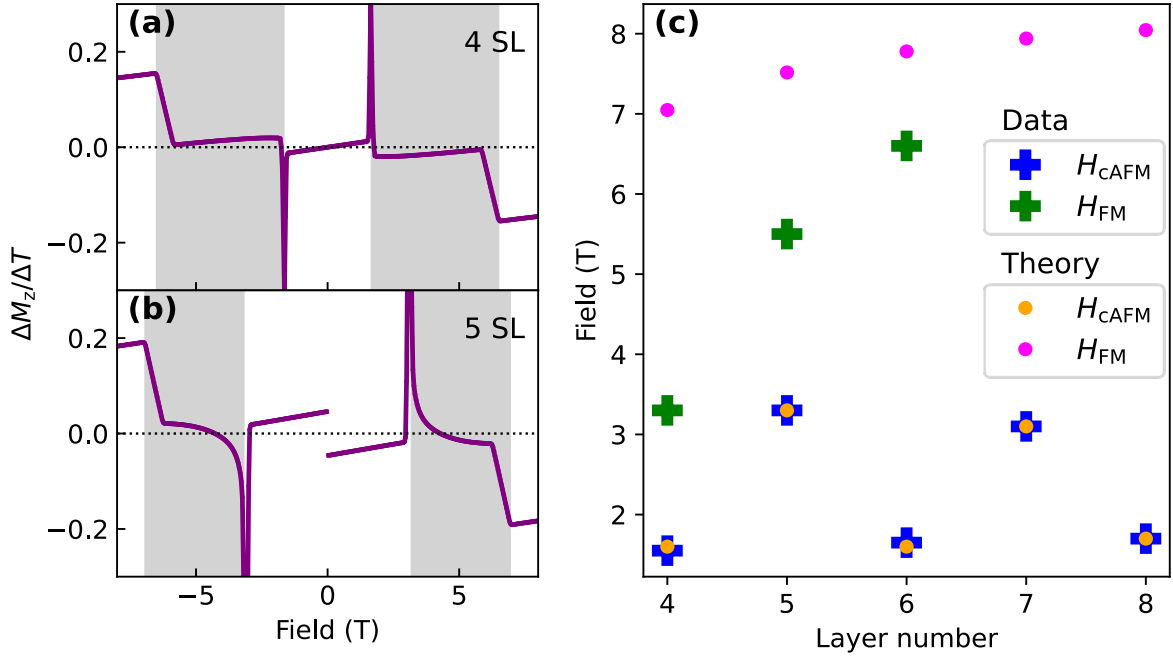

**Figure 2.** (Color online) **Theoretical calculations of $\Delta M_z/\Delta T$ and transition fields.** The transition between magnetic states and the temperature dependent magnetization is calculate using our theoretical model described in Supplementary Note 3, with exchange and anisotropy parameters $J = 2.3$ T and $K = 0.35\ J$, and theoretical temperature values $T = 2$ and $T = 4$ for the $\Delta M_z/\Delta T$ plots. The difference in z-direction magnetization between the two temperatures is shown for 4- and 5-layer samples in **a** and **b**. In panel **c**, the AFM-cAFM transition fields ($H_{cAFM}$) and cAFM-FM transition fields ($H_{FM}$) extracted from the real data (crosses) are compared to the same transition fields calculated from the theory (dots). The comparison reveals a dramatic difference in the cAFM-FM transitions, especially for the thinnest samples.

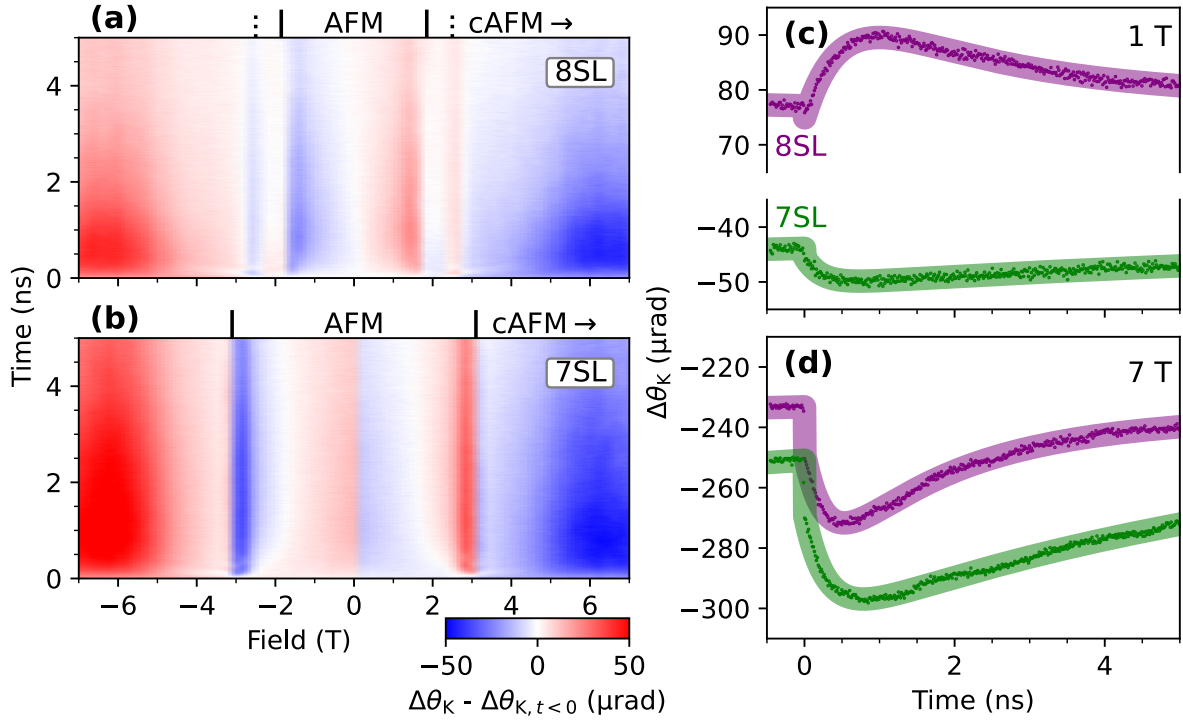

**Figure 3.** (Color online) **Field dependence of time-resolved Kerr measurements.** We show the time-resolved data after subtracting the background signal across the full field range (swept from negative to positive) for 8 SL (**a**) and 7 SL (**b**) sample regions. The AFM-cAFM transitions are indicated with solid lines, and a secondary transition in the 8 SL sample between different cAFM configurations is indicated with a dotted line. Individual signals at 1 T and 7 T, where both samples are in the AFM and cAFM states respectively, are plotted in **c** and **d**. In the cAFM state, both regions have similar signals representing a decrease in $M_z$ from heating. In the AFM state, $dM_z/dT$ becomes positive for an even-layered sample, giving a signal of opposite sign, whereas for the odd-layered sample it remains negative. The initial step-like component is also smaller, and remains negative in the even-layered sample, in contrast to the other components.

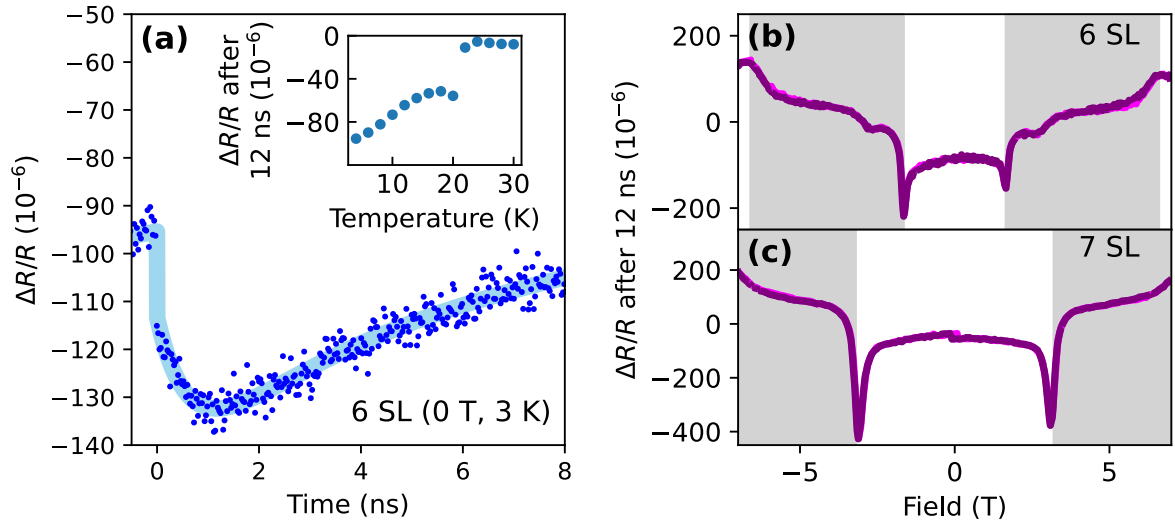

**Figure 4.** (Color online) **Magnetic state dependence of time-resolved reflectivity.** The time-resolved reflectivity (shown for a 6-layer region of the sample at 3 K with no applied field in **a**) shows the same general features seen in the Kerr measurements. Note that this large signal vanishes above $T_N$ (inset in **a**), and the background signals (**b** and **c** for 6- and 7-layer samples respectively) show clear cusps near the transition points (where, as before, the cAFM region is indicated by grey shading), indicating it is genuinely connected to the magnetic state. It differs notably in that the signal is (nearly) symmetric about the field, compared to the antisymmetric Kerr. Additionally, the even-layer sample has a large signal with zero applied field, where both the time-resolved Kerr signal and the expected change in net magnetization with temperature from calculations are very small.

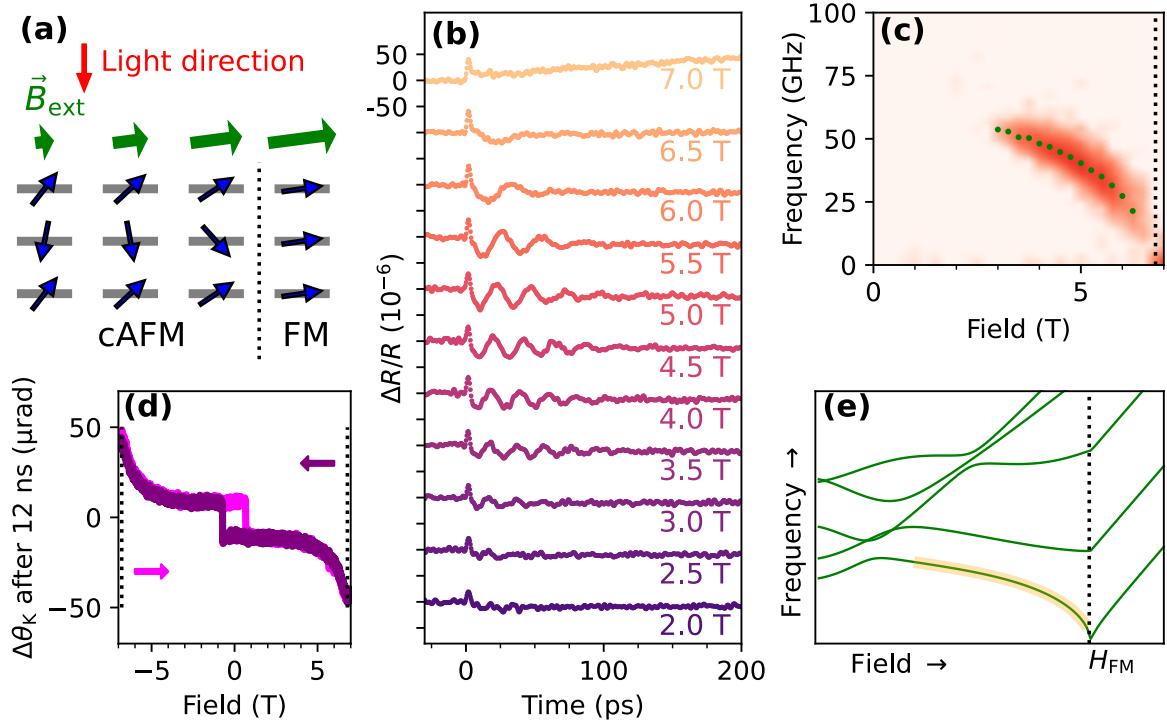

**Figure 5.** (Color online) **Magnon oscillations in 5-SL MnBi$_2$Te$_4$.** The external field is applied at a small (≈ 7.5°) angle with respect to the sample plane, as shown in **a**, which causes the spins to begin tilting for all non-zero field strengths, eventually transitioning to (nearly in-plane) FM order at high field. Transient reflectivity measurements at varying fields are shown in **b** (curves offset for clarity), where starting at around 3 T we can observe a distinct oscillation which decreases in frequency with increasing applied field before vanishing close to 7 T. The field dependence of this mode can be seen more clearly in the FFT amplitude (**c**, central frequencies from fitting shown with green dots), dropping to zero frequency at ~6.8 T (dashed line). Like the out-of-plane field case, we observe a long-lived component of the $\Delta\theta_K$ signal, which can be measured as a function of field (**d**, with arrows indicating sweep direction) to reveal the magnetic state. A cusp at 6.8 T (dotted line, the same field where the magnons vanished) can be observed, which we attribute to the cAFM-FM transition. Comparing our observations to magnon modes calculated from theory (green lines in **e**), they appear to match well with the portion of the lowest-frequency mode just below the cAFM-FM transition field (yellow highlight in **e**).

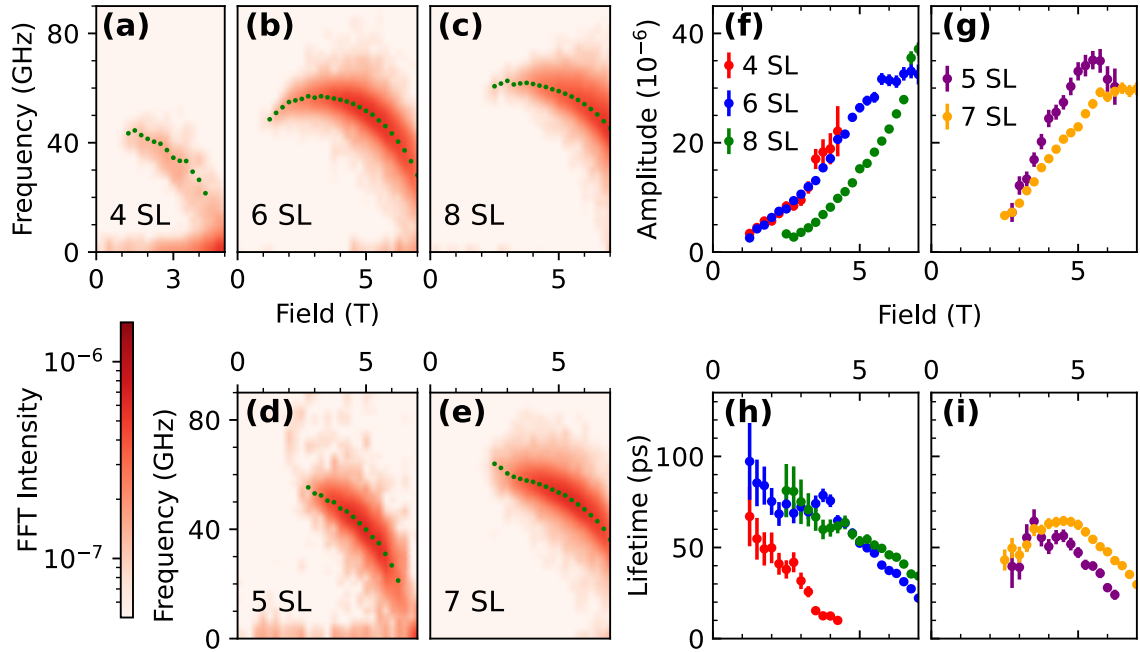

**Figure 6.** (Color online) **Field- and thickness-dependence of magnons in MnBi$_2$Te$_4$.** In **a-e** we show the FFT amplitude of the transient reflectivity for various thicknesses (all at 4 K) along with central frequencies from fits (green dots), where a distinct difference can be observed between the even (**a-c**) and odd (**d,e**) layered sample regions. We further plot how the amplitude (**f,g**) and lifetime (**h,i**) of the magnon mode, obtained from least-squares fitting, vary across magnetic field for each of the sample regions, where again distinct differences can be observed between the even (**f,h**) and odd (**g,i**) layered regions.